\documentstyle[epsfig]{elsart}
\begin{document}
\hyphenation{a-na-ly-sis com-po-si-tion e-ner-gy showers ex-pe-ri-men-tal
pro-ba-bi-li-ties si-mu-la-tions con-si-de-red ge-ne-ra-tors mul-ti-pli-ci-ty
re-gis-te-red po-pu-lar a-na-ly-ses de-ve-lop-ment me-tho-di-ca-lly
par-ti-cu-lar}
\begin{frontmatter}
\title{
The information from muon arrival time distributions of high-energy EAS as 
measured with the KASCADE detector
}
\begin{center}
KASCADE Collaboration
\end{center}
\author[bbb]{T.~Antoni},
\author[aaa]{W.D.~Apel},
\author[bbb]{A.F.~Badea}$^,$\footnote{corresponding author: 
Florin.Badea@ik.fzk.de}$^,$\footnote{on leave of absence from National 
Institute of Physics and Nuclear Engineering, Bucharest, Romania},
\author[aaa]{K.~Bekk},
\author[aaa]{A.~Bercuci}$^{,\,\,2}$,
\author[aaa,bbb]{H.~Bl\"umer},
\author[aaa]{H.~Bozdog},
\author[ccc]{I.M.~Brancus},
\author[aaa]{C.~B\"uttner},
\author[ddd]{A.A.~Chilingarian},
\author[bbb]{K.~Daumiller},
\author[aaa]{P.~Doll},
\author[aaa]{J.~Engler},
\author[bbb]{F.~Fessler}, 
\author[aaa]{H.J.~Gils},
\author[bbb]{R.~Glasstetter},
\author[bbb]{R.~Haeusler},
\author[aaa]{A.~Haungs},
\author[aaa]{D.~Heck},
\author[bbb]{J.R.~H\"orandel},
\author[bbb]{A.~Iwan}$^,$\footnote{on leave of absence from University of Lodz,
Lodz, Poland},
\author[bbb,aaa]{K.H.~Kampert},
\author[aaa]{H.O.~Klages},
\author[aaa]{G.~Maier},
\author[aaa]{H.J.~Mathes},
\author[aaa]{H.J.~Mayer},
\author[bbb]{J.~Milke},
\author[aaa]{M.~M\"uller},
\author[aaa]{R.~Obenland},
\author[aaa]{J.~Oehlschl\"ager},
\author[bbb]{S.~Ostapchenko}$^,$\footnote{on leave of absence from Moscow State 
University, Moscow, Russia},
\author[ccc]{M.~Petcu},
\author[aaa]{H.~Rebel},
\author[aaa]{M.~Risse},
\author[aaa]{M.~Roth},
\author[aaa]{G.~Schatz},
\author[aaa]{H.~Schieler},
\author[aaa]{J.~Scholz},
\author[aaa]{T.~Thouw},
\author[bbb]{H.~Ulrich},
\author[bbb]{J.H.~Weber},
\author[aaa]{A.~Weindl},
\author[aaa]{J.~Wentz},
\author[aaa]{J.~Wochele},
\author[eee]{J.~Zabierowski}

\address[bbb]{University of Karlsruhe, Institut f\"ur Experimentelle
Kernphysik, 76021 Karlsruhe, Germany}
\address[aaa]{Forschungszentrum Karlsruhe, Institut f\"ur Kernphysik,
76021 Karlsruhe, Germany}
\address[ccc]{National Institute of Physics and Nuclear Engineering, 7690 
Bucharest, Romania}
\address[ddd]{Cosmic Ray Division, Yerevan Physics Institute, Yerevan 36,
Armenia}
\address[eee]{Soltan Institute for Nuclear Studies, 90950 Lodz, Poland}

\begin{abstract} 
Using the facilities of the KASCADE Central Detector EAS muon arrival time 
distributions, observed with reference to the arrival time of the first 
locally registered muon, and their correlations with other EAS observables 
have been experimentally 
\newpage
\noindent
investigated. The variation of adequately defined time parameters with the 
distance $R_\mu$ from the EAS axis has been measured. The experimental data 
enable a study of
the sensitivity of such local arrival time distributions, which characterise 
the structure of the shower disc, to the mass composition of cosmic rays in the 
energy region around the knee. For that purpose, 
nonparametric multivariate even-by-event analyses have been performed for an 
estimate of the mass composition specified by three different mass groups,
invoking detailed Monte Carlo simulations of the EAS development. It 
turns out that local muon arrival time distributions, without information 
on the  curvature of the shower disc, display a minor sensitivity to the mass 
of the EAS inducing particle, at least for distances from the shower axis 
\mbox{$R_\mu<100$ m}. 
The measurements comprise a subset of all EAS events registered by KASCADE
due to the observation conditions of the arrival time distributions, with a 
threshold of the muon energy \mbox{$E_{th}=2.4$ GeV} and a minimum multiplicity 
$n_{th}$ for being accepted in the observed data samples. This subset is 
sensitive to variations of the integral EAS muon energy spectrum.
By studying the event acceptance in the registered samples on basis of 
Monte Carlo simulations a test of the consistency of the Monte Carlo simulations
with the data is enabled, comparing the results inferred from observations at 
different $R_\mu$ and different $n_{th}$ values. 
Within the present uncertainties the results of such a test show a remarkable 
agreement of the experimental findings with the Monte Carlo simulations, using 
the QGSJET model as generator of the high-energy hadronic interactions. 
\end{abstract}
\end{frontmatter}
\section{Introduction}
The time delay of the particles inside the front of Extensive Air Showers (EAS) 
and the temporal structure of different EAS components are a subject of 
longstanding interest of cosmic ray research. In fact the first experimental 
studies have been performed in 1953 by Bassi, Clark and Rossi \cite{basi53} and 
Jelly and Whitehouse \cite{jell53}, followed by many others 
\cite{lins61}-\cite{good79}. The renewed interest arises from recent 
measurements using advanced detector facilities like the COVER –PLASTEX 
detector within the GREX array \cite{agne97}-\cite{ambr99}, the EAS TOP array 
\cite{eas93} or the  facilities of the KASCADE experiment \cite{klag97}. 
Investigations of the structure of the EAS disc by the KASCADE experiment are 
focused in particular on the muon component \cite{rebe95}-\cite{1anto01}.
Due to the reduced influence of multiple Coulomb scattering of GeV muons and 
the absence of absorption, muons travel nearly in straight lines from the 
locus of production to the observing detector. Thus muon arrival time 
distributions, observed at large distances from the shower axis,
are expected to map the longitudinal EAS development and to reflect the 
distribution of production heights via the time-of-flight of the muons. 
With some simplifications, using a triangulation procedure, the distribution 
of the production heights could be estimated from the time delay of the 
observed muons relative to the arrival time of the shower centre 
\cite{rebe95,dani95,bran97}. Basically the same information could be
alternatively deduced from the distribution of angles of muon incidence 
relative to the shower axis \cite{bern96}. 
Corresponding measurements are also a current subject of  KASCADE 
investigations using the Muon Tracking Device \cite{zabi01}. Angle-of-incidence 
observations or the combination with arrival time measurements (as discussed 
with the term Time Track Complementary~\cite{dani95,2ambr97}) do not 
reveal additional basic information \cite{bran97} though they provide an 
interesting practical alternative.\\
\\
The basic sensitivity of muon arrival time distributions (whose phenomenological
features seen with the KASCADE experiment are reported in Ref.~\cite{bran99}) 
to longitudinal EAS development and to the elongation rate 
\cite{lins81,walk81,1bade01} puts the question which particular
parameters of the observed distributions provide useful signatures of the mass 
of the EAS primary and under which conditions they are helpful for the 
determination of the mass composition of the primary cosmic rays. The present 
paper addresses experimentally the question of the sensitivity to the primary 
mass on the basis of results from the KASCADE experiment, in particular of EAS 
event-by-event measurements of the temporal EAS structure (shower thickness) 
observed at relatively small distances to the shower axis 
\mbox{$R_\mu < 100$ m} for the primary energy range around the knee. 
Methodically non-parametric statistical analysis techniques of multivariate 
distributions 
\cite{chil89} are applied for the mass classification of the observed EAS by 
their time parameters, derived from the muon arrival time distributions, and by 
their correlations with other EAS observables (like the shower size $N_e$ and 
the muon content $N_\mu$). For muon arrival time distributions the methods are
outlined in Refs.~\cite{rebe95,bran97}. These procedures 
avoid the bias of pre-chosen functional forms and compare to reference 
patterns, derived from extensive Monte Carlo simulations, by use of the EAS 
Monte Carlo simulation code CORSIKA \cite{heck31}, thus including 
all natural EAS fluctuations. By applying additional detailed detector 
simulations based on GEANT \cite{brun87}, the distortions arising from the 
experimental conditions \cite{haeu01} are taken into account.\\
The analysis of the EAS observations in terms of the mass of the primary 
introduces necessarily some model dependence by the high-energy hadronic 
interaction models invoked as generators of the Monte Carlo simulations. 
Therefore the results are subject of the uncertainities of the hadronic 
interaction model, used for the analysis. A possible way to evaluate the 
quality of a model is to derive the primary mass composition from the analysis 
of different observables and to consider the agreement or disagreement of 
the results. This idea has been worked out in a consistent and efficient 
manner \cite{roth01} with an extended set of observables of the KASCADE 
experiment, showing that there exist systematic differences in the estimate 
of $\langle lnA \rangle$, e.g., derived from different combinations of 
(correlated) observables \cite{2anto01}, revealing the limitations 
of popular models like QGSJET \cite{kalm97}, VENUS \cite{wern93} 
and SIBYLL \cite{flet94}. This indicates that improvements of the models are 
urgently needed on basis of experimental findings. The QGSJET model is 
actually considered to be that one of the best internal consistency 
\cite{anto99}. Hence we refer furtheron to the QGSJET model.\\
The muon arrival time measurements imply a particular selection of EAS, with a 
distortion of the original mass composition in the subset of classified EAS.
This is due to the energy threshold $E_{th}$ (\mbox{2.4 GeV} for KASCADE) of 
the muon detection and the condition of a minimum multiplicity $n_{th}$ ($=3$ 
muons detected by the KASCADE Central Detector) for reconstructing an arrival 
time distribution. The distortion is dependent on the particular values of 
$E_{th}$, $n_{th}$, on the mass $A$ of the primary and the muon content $N_\mu$,
but also on the radial distance $R_\mu$ of the arrival time 
measuring detector from the shower axis. In order to restore the original mass 
composition corresponding efficiency (acceptance) correction factors have to be 
applied to the identified mass groups. These factors can be only determined by 
Monte Carlo simulations, studying the cuts implied by the experiment and their 
effect on the efficiency. 
An efficient and sensitive test of the interaction model used for the 
simulations and of the particle tracking procedures is enabled by applying the 
calculated correction factors to the results found at {\it different} $R_\mu$ 
and to look for the consistency of the resulting mass composition. 
Such a test of the Monte Carlo simulations and of the QGSJET model, 
especially in view of predictions of the muon energy spectra and muon densities 
and their fluctuations in event-by-event observations, is an important aspect 
of muon arrival time studies of this paper. The results add to the conclusions 
about the difficulties of the model in interpreting consistently muon density 
measurements with {\it different} muon {\it energy} thresholds \cite{3anto01}. 

\section{Muon arrival time measurements}
Measured arrival times $\tau_{1\mu} < \tau_{2\mu} < \tau_{3\mu} <...$ 
of muons registered by timing detectors at a distance $R_\mu$ from 
the shower axis, corrected by $\pm R_\mu \tan\theta /c $ ($c$ - speed of
light) in order to eliminate the distortions due to the shower inclination,
refer to an experimentally provided zero-time. Depending on the choice of the 
kind of zero-time reference, there are two 
different types of muon arrival time distributions. By the use of the arrival 
time $\tau_{c}$ of the shower core as reference global arrival times are 
observed \cite{haeu01}: 
\begin{eqnarray}
\Delta\tau^{glob}_i(R_{\mu})=\tau_{i\mu}(R_{\mu})-\tau_{c}\,\,,
\end{eqnarray}
e.g. for the foremost muon registered at $R_\mu$:
\begin{eqnarray}
\Delta\tau_{1}^{glob}(R_{\mu})=\tau_{1\mu}(R_{\mu})-\tau_{c}\,\,.
\end{eqnarray}
This type of time distributions informs about the curvature of the shower 
front as well as about the structure of the disc. The  arrival time of the
shower core is difficult to determine with sufficient precision. Therefore the 
analysis has been restricted to "local" times, which refer to the arrival 
of the foremost muon locally registered by the detector:
\begin{eqnarray}
\Delta\tau^{loc}_i(R_{\mu}) = \tau_{i\mu}(R_{\mu})-\tau_{1\mu}(R_{\mu})\,\,.
\end{eqnarray}
(with omitting further the label loc), informing only about the thickness and 
the structure of the muon disc. Implications of observations of local 
muon arrival time distributions due to the fluctuations of the arrival of the 
first registered muon have been discussed in Ref.~\cite{haeu01}. 
For event-by-event observations with a fluctuating number of muons 
(multiplicity), the individual relative arrival time distributions can be 
characterised by the mean values $\Delta\tau_{mean}$, and by various quantiles 
$\Delta\tau_{q}$, like the median $\Delta\tau_{0.50}$, the first quartile 
$\Delta\tau_{0.25}$ and the third quartile $\Delta\tau_{0.75}$ 
(for details of the definition see Refs.~\cite{bran98,bran99,1anto01}). Their 
mean values and dispersion (standard deviations) represent the time profile of 
the EAS muon component.\\
\\
KASCADE, whose layout is described in more detail in Ref.~\cite{klag97}, is a 
multidetector system, installed in Forschungszentrum Karlsruhe (\mbox{110 m} 
a.s.l.), 
Germany, for the observation of extensive air showers in the primary energy 
range around the {\it knee}. One part is an array of 252 detector stations, 
distributed over an area of \mbox{$200 \times 200$ m$^2$} on a grid of 
\mbox{13 m} spacing for measuring the electron-photon component and the muon 
component with a threshold of \mbox{5 MeV} and \mbox{230 MeV}, respectively 
and providing the basic information about arrival direction, core position, 
electron and muon sizes of the observed EAS. In particular, from 
the data of the field array the so-called truncated muon number $N_\mu^{tr}$, 
i.e. the muon density integrated between \mbox{40 m} and \mbox{200 m}, is 
derived and used in the KASCADE case (between $10^{14}$ and \mbox{$10^{16}$ eV})
as an approximate mass independent energy identifier \cite{webe97}. The 
location of the shower core can be determined (inside the fiducial area) with 
an uncertainty of less than \mbox{3 m}. The arrival direction of the shower is 
reconstructed with an uncertainty better than $0.5^\circ$. Details of the 
reconstruction procedures are described in Ref.~\cite{4anto01}. As additional 
muon detector Limited Streamer Tubes tracking detectors have been installed in 
an underground tunnel for measurements of the lateral distribution 
(\mbox{$E_{th} = 0.8$ GeV}) and of muon angles-of-incidence 
distributions \cite{zabi01}.\\
\\
The muon arrival time measurements use, in particular, the facilities of the 
Central Detector \cite{engl99} of KASCADE. It 
is basically an iron sampling calorimeter (with an area of 
\mbox{$16 \times 20$ m$^2$}, set up with liquid TMS and TMP ionisation chambers)
for 
the identification and energy measurement of hadrons. In the basement of the 
set-up, below \mbox{3800 t} of iron and concrete, large-area position-sensitive 
multiwire proportional chambers (MWPC) \cite{bozd01} are operated for the 
identification of muons with \mbox{2.4 GeV} energy threshold. The performance 
of the detection system of the MWPC is improved by a layer of streamer tubes 
\cite{5anto01}. The trigger plane of the calorimeter is a system of 456 plastic 
scintillation detector elements (\mbox{$47.5 \times 47.5 \times 3$ cm$^3$} in 
size, each separated by a wavelength shifter) for providing a fast 
trigger signal (in addition to the trigger from the field array) for the MWPC, 
and for the timing measurements. Fast electronics \cite{bren98} records 
low (muons) and high energy deposits (cascading hadrons). In order to 
remove the signals from cascading hadrons an upper limit of the energy deposit 
of \mbox{30 MeV} in each of the 456 scintillation counters is imposed. 
An amount of 24 millions of KASCADE events are used for the
analysis \cite{2bade01}. Muon arrival time distributions 
have been reconstructed for muons with \mbox{$E_{th}=2.4$ GeV} and for events 
with a number $n \ge 3$ of registered  muons;
after applying also some general cuts concerning the core position (within 
\mbox{100 m} from the array centre), the angle of EAS incidence ($< 40^\circ$) 
and $log_{10}N_\mu^{tr}$ ($> 3.6$) the sample shrinked to approx. 240~000 
showers. The phenomenological 
features of the observed muon arrival time distributions have already been
reported in Ref.~\cite{1anto01}, where further experimental details are 
communicated.

\section{EAS simulations}
The interpretation of the measured muon arrival time distributions and their 
correlations with other shower parameters need a-priori knowledge to be deduced 
from Monte Carlo simulations of the EAS development. The present analysis is 
based on simulations with the code CORSIKA (version 5.62) with a full and 
detailed simulation of the detector response. The simulations use the QGSJET 
(version 1998) model \cite{kalm97} as generator for high-energy interactions and
GHEISHA \cite{fese85} for interactions below \mbox{$E_{lab}= 80$ GeV}. The 
electromagnetic part is treated by the EGS4 program \cite{nels85}. 
Earth magnetic field, observation level and particle detection thresholds have 
been chosen in accordance to the experimental situation. The U.S. standard 
atmosphere \cite{heck31} has been adopted for the simulations. The simulations 
have been performed  for three different classes of primaries: protons (H) for 
the light group, oxygen (O) for the CNO-group and iron (Fe) for the heavy group.
The energy range covered by the simulations extends from 
\mbox{$5.0\cdot10^{14}$ eV} to \mbox{$3.06\cdot10^{16}$ eV} and the zenith 
angles comprise the range of $0^\circ \leq \theta \leq 40^\circ$. The centres 
of the simulated showers have been positioned inside a quadratic area of
\mbox{$210 \times 210$ m$^2$}, slightly extending the KASCADE area. 
For each primary type approximately 90000 showers have been simulated, with 
decreasing numbers of shower events for higher primary energies and the larger 
zenith 
angles, due to restrictions in the computing time. Finally a sample for 
$0^\circ \leq \theta \leq 24^\circ$ (corresponding to a third of the considered
$\sec\theta$ - range of $(1,\sec 40^\circ)$) was used in the analysis. 
The simulated primaries (separately for each of the cases H, O, Fe) have been 
weighted according to: 
\begin{eqnarray} 
dN=const. \cdot E_0^{-\gamma}\cdot  \sin\theta \cdot \cos\theta \cdot 
D_{core} \cdot dE_0 \cdot d\theta \cdot dD_{core}
\label{ecudn} 
\end{eqnarray}
where $dN$ is the number of primaries with energy between $E_0,E_0+dE_0$,  
zenith angle between $\theta,\theta+d\theta$, intersecting the plane of 
the detector array between $D_{core},D_{core}+dD_{core}$. 
A fixed spectral index $\gamma = 2.7$ over the whole primary energy 
\begin{figure}[t]
\begin{center}
\epsfig{file=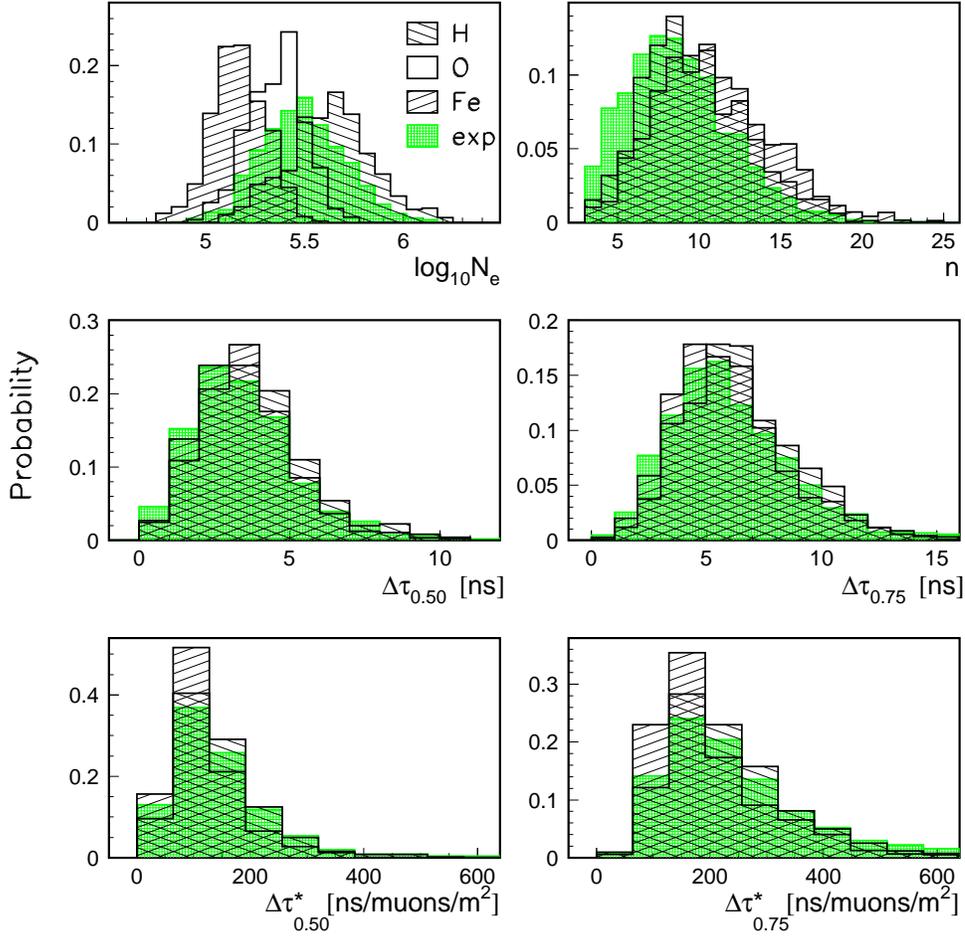, width=13.0cm}
\caption{The distribution of various shower observables of the sample, prepared 
by simulations for different types of primaries, compared with the distribution 
of the sample of experimental data for $4.05 < log_{10}N_{\mu}^{tr} \leq 4.28$,
\mbox{80 m} $< R_\mu \leq$ \mbox{90 m} and $0^\circ \leq \theta \leq 24^\circ$.
The quantity $n$ is the multiplicity of muons (\mbox{$E_{th}=$2.4 GeV}) detected
with the MWPC setup of the Central Detector.
}
\label{fig1}
\end{center}
\end{figure}
range is adopted for all primaries (H, O, Fe).
More details about the simulations are given in Refs. \cite{haeu00} and 
\cite{2bade01}. As input for the further analysis  of the muon arrival time 
distributions and their comparison with model predictions the following shower 
parameters and their correlations are regarded: 
\begin{itemize}
\item
the shower size $N_e$;
\item
the truncated muon number $N_{\mu}^{tr}$;
\item
the multiplicity $n$ of muons (\mbox{$E_{th}=2.4$ GeV}) detected in the 
facilities of the Central Detector;
\item
the quantiles $\Delta\tau_q$ of the local muon (\mbox{$E_{th}=2.4$ GeV}) 
arrival time distributions at various distances $R_\mu$ from the shower centre;
\item
the reduced quantiles $\Delta\tau_q^*=\Delta\tau_q/\rho_\mu$ of the local muon 
arrival time distributions (\mbox{$E_{th}=2.4$ GeV}), i.e. the quantiles 
divided by the density $\rho_\mu$, where  $\rho_\mu$ is estimated from the 
observed multiplicity and the effective area of the muon detector set-up.
\end{itemize} 
Fig.~\ref{fig1} displays  frequency distributions of some shower parameters, 
prepared by the simulation for different primaries and compared to the actual 
experimental observations for particular $log_{10}N_{\mu}^{tr}$
and $R_\mu$ bins. Each distribution is normalised (only for the 
presentation of $N_e$ the distribution of oxygen-induced EAS is additionally 
shown). 
There are some discrepancies in the multiplicity distribution between 
simulations and experimental observations. They may arise from an imperfect
adoption of the primary energy spectrum and from efficiency effects not fully
corrected for. Since they also affect the fluctuations of the local time 
parameters originating from the multiplicity dependence (discussed in detail 
in Ref.~\cite{haeu01}), this analysis uses preferentially the reduced quantities
$\Delta\tau_q^*$. The fluctuations largely cancel in the reduced parameters. 
Actually it has been argued that the reduced quantiles $\Delta\tau_q^*$ could 
exhibit enhanced mass sensitivity since they include approximately the  
$\Delta\tau_q(R_\mu) - \rho_\mu (R_\mu)$ correlation \cite{bran01}. However 
this sensitivity is obscured by the limitations of the experimental response.
CORSIKA simulations show \cite{bran97,bran01} that the age parameter is 
carrying relevant information about the longitudinal development. This aspect
has not been worked out by the present analysis.

\section{Non-parametric multivariate analyses}
Non-parametric statistical methods are applied in studies of multidimensional 
distributions of observables allocating the single observed events to different 
classes (in our case to proton, oxygen, or iron primaries) by comparing the 
observed events with the simulated distributions without using a pre-chosen 
parameterisation. The methods of decision making and procedures of applications 
to cosmic ray data analyses are extensively described in 
Refs.~\cite{chil89,2anto01} and outlined  for analyses of muon arrival time 
distributions in Refs.~\cite{rebe95,bran97}. The procedures take 
into account the effects of the EAS fluctuations in a quite natural way and are 
able to specify the uncertainties, by an estimate of the true-classification 
and misclassification probabilities. The classification probabilities are 
determined by the extent to which the likelihood functions of the individual 
classes, derived from the simulations, are overlapping. Basically such pattern 
recognition methods, using trained neural networks or Bayes decision rules, 
depend on the hadronic model generating the reference patterns for the 
experimental observables. 
The concept of the present analysis is to determine in a first step the 
sensitivity of various observable correlations to the mass of the primary, by
applying the one-leave-out test to the sample of distributions prepared by Monte
Carlo simulations (QGSJET model). The one-leave-out test determines the
probability that a multidimensional event, taken from the considered (simulated)
distribution, will be correctly ("true") or incorrectly ("false") classified by
the pattern recognition procedure (see appendix of Ref.~\cite{2anto01}).
With respect to the experimental data sample the 
studies are performed not only for six different $log_{10}N_{\mu}^{tr}$ ranges  
between 3.60 and 5.00 (covering the knee-range) but also for four different 
$R_\mu$-ranges (\mbox{45-65 m}, \mbox{65-80 m}, \mbox{80-90 m}, 
\mbox{90-100 m}). This allows to explore trends of the classification and 
misclassification probabilities and other quantities with varying $R_\mu$.
Using the determined classification probabilities the mass composition is 
reconstructed from the data samples measured at different $R_\mu$-ranges. The 
resulting mass composition generally differs from the primary mass composition
searched for, since the analysed data samples represent selections in 
the multidimensional space of all EAS observables. Hence efficiency correction 
factors have to be determined, whose application should lead to a primary mass 
composition consistent for all $R_\mu$-ranges. The determination of the 
efficiency correction factors invokes again Monte Carlo simulations, and 
the extent of agreement of the primary mass composition resulting from 
different $R_\mu$ ranges provides a test for the procedures and 
the generator of the Monte Carlo simulations.

\subsection{The classification and misclassification  probabilities}
\begin{figure}[p]
\begin{center}
\epsfig{file=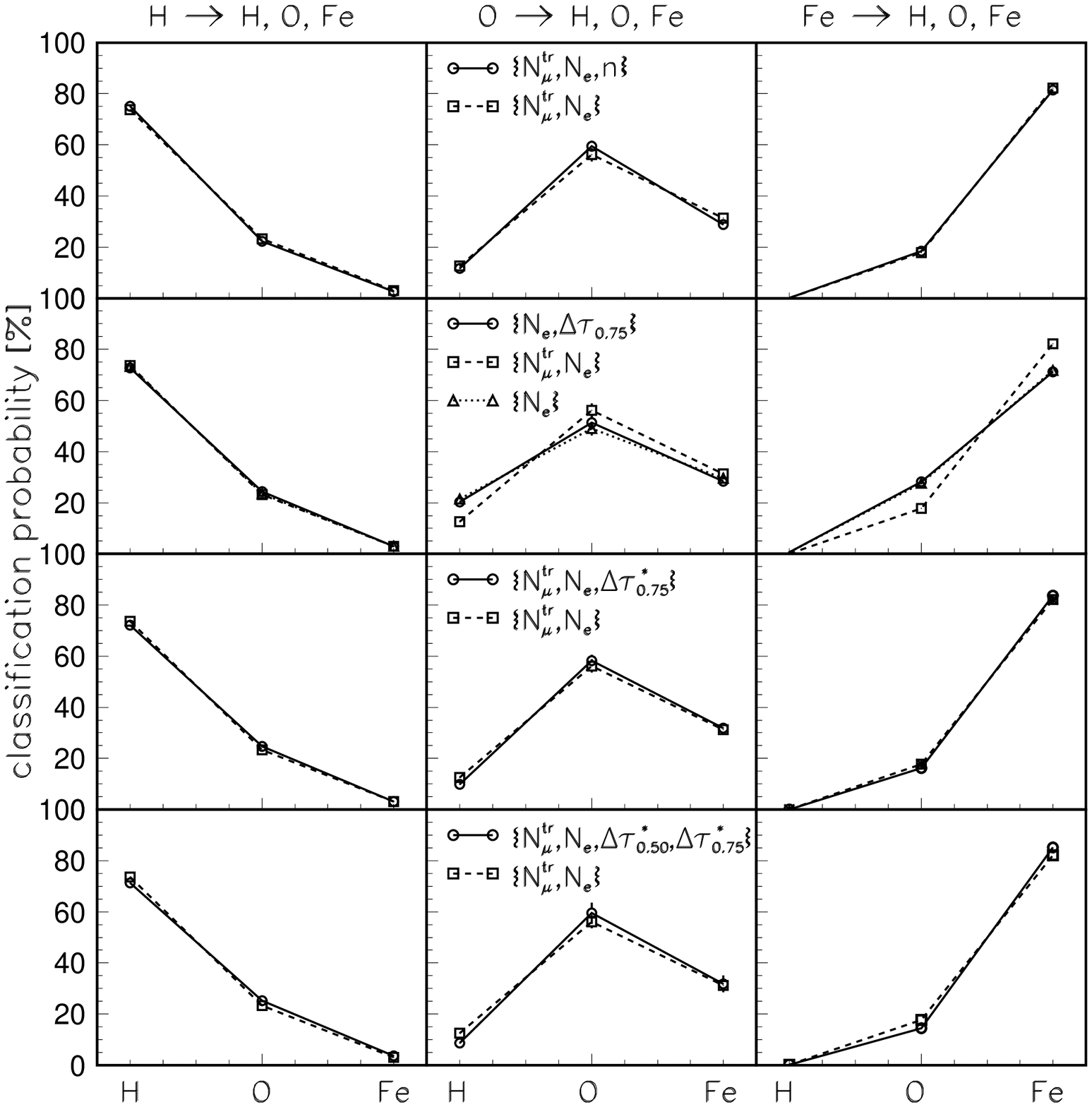, width=14.0cm}
\vspace*{-0.6cm}
\caption{Classification and misclassification probabilities determined by Bayes 
decision making for three different classes (H, O, Fe) and for distribution 
of different types of correlated EAS observables with 
$3.83 < log_{10}N_{\mu}^{tr} \leq 4.05$,
\mbox{80 m} $< R_\mu \leq$ \mbox{90 m}, $0^\circ \leq \theta \leq 24^\circ$. The
results of different observable combinations are connected by (full and dashed)
lines.
}
\label{fig2}
\vspace*{0.4cm}
\epsfig{file=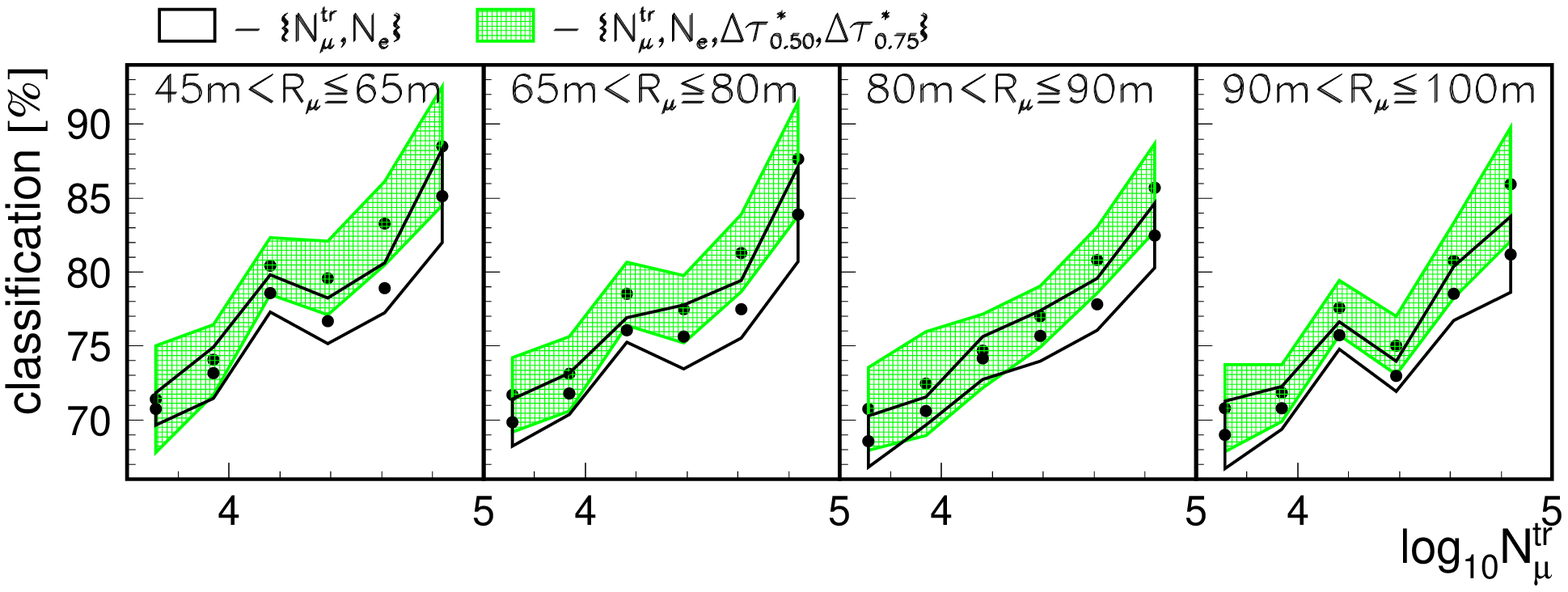, width=14.0cm}
\vspace*{-0.8cm}
\caption{The dependence of the averaged true-classification probability 
from $\{N_{\mu}^{tr},N_e\}$ and $\{N_{\mu}^{tr},N_e,\Delta\tau_{0.50}^*,
\Delta\tau_{0.75}^*\}$ correlations ($0^\circ \leq \theta \leq 24^\circ$).
}
\label{fig3}
\end{center}
\end{figure}

As example in Fig.~\ref{fig2} the true-classification and misclassification 
probabilities, inferred with the one-leave-out test from the simulation sample, 
are displayed for H, O, Fe primaries and for different combinations of 
observables of EAS events. The events have been registered with the conditions 
of muon 
arrival time measurements in a particular $\{ log_{10}N_{\mu}^{tr}, R_\mu\}$
- range. With little surprise it is noted that the mass discrimination is 
dominated by the $\{N_{\mu}^{tr},N_e\}$ correlation (see Fig.~\ref{fig3}), and 
it is also evident 
that local time parameters (shown for the median $\Delta\tau_{0.50}$ or 
the third quartile $\Delta\tau_{0.75}$ and their reduced values
$\Delta\tau_{0.50}^*$, $\Delta\tau_{0.75}^*$) have minor influence on the 
discrimination. We emphasise that this statement holds for {\it local} arrival 
time distributions in the studied range of relatively small distances $R_\mu$ 
(\mbox{$<$ 100 m}) from the shower centre. Theoretical studies
\cite{2bade01,bran01} indicate more pronounced differences in the 
temporal structure of the shower disc with increasing $R_\mu$ and
$log_{10}N_{\mu}^{tr}$. In contrast to results ignoring the detector 
response \cite{bran01} the use of reduced parameters 
$\Delta\tau_{0.50}^*$, $\Delta\tau_{0.75}^*$ does not significantly improve the 
mass discrimination. Fig.~\ref{fig3} shows the true classification probabilities
(and uncertainties) averaged over all classes with increasing 
$log_{10}N_{\mu}^{tr}$ indicating a marginal systematic 
improvement if the time information is added. The uncertainties are determined 
by the bootstrap method \cite{chil89}, which basically consists of applying 
the classification procedure for a test sample several times, 
thus deriving an average value and the variance.

\subsection{Reconstruction of the mass composition from the observed data 
samples}
The true-classification $P_{i \rightarrow i}$ and misclassification
\begin{figure}[b]
\begin{center}
\epsfig{file=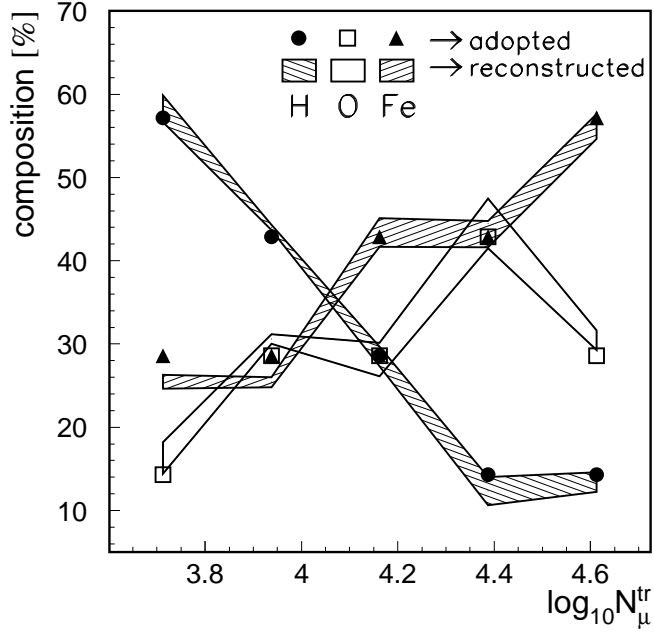, width=9.0cm}
\caption{Test of the reconstruction of  an arbitrarily adopted mass composition 
of the event samples on basis of the $\{N_{\mu}^{tr},N_e,\Delta\tau_{0.50}^*,
\Delta\tau_{0.75}^*\}$ correlation, observed at \mbox{80 m} $< R_\mu \leq$ 
\mbox{90 m}, $0^\circ \leq \theta \leq 24^\circ$.}
\label{fig4}
\end{center}
\end{figure} 
$P_{i \rightarrow j}$ probabilities, deduced for all studied 
$\{log_{10}N_{\mu}^{tr}, R_\mu\}$ ranges, are used for the reconstruction 
\cite{bran97} of the mass composition of the samples of registered events by 
inverting the system of linear equations:
\begin{eqnarray}
\begin{array}{l}
n_H^\prime = n_H\cdot P_{H \rightarrow H} + n_O\cdot P_{O \rightarrow H} + 
n_{Fe}\cdot P_{Fe \rightarrow H}\\
n_O^\prime = n_H\cdot P_{H \rightarrow O} + n_O\cdot P_{O \rightarrow O} + 
n_{Fe}\cdot P_{Fe \rightarrow O}\\
n_{Fe}^\prime = n_H\cdot P_{H \rightarrow Fe} + n_O\cdot P_{O \rightarrow Fe} + 
n_{Fe}\cdot P_{Fe \rightarrow Fe}
\end{array}
\end{eqnarray}
where $n_H$, $n_O$, $n_{Fe}$ are the true numbers, defining the mass 
composition in the sample $N=n_H+n_O+n_{Fe}$, getting altered to 
$n_H^\prime$, $n_O^\prime$, $n_{Fe}^\prime$ because of the misclassifications.
In Fig.~\ref{fig4} the reconstruction of an arbitrarily adopted mass 
composition of the event samples (displayed by the corresponding symbols), is 
shown with the resulting uncertainties as varying with $log_{10}N_{\mu}^{tr}$.
The result of the application of the reconstruction procedures to experimental 
samples measured with KASCADE, observed at different distances $R_\mu$ from 
the shower centre, are shown in the upper panels of Fig.~\ref{fig5}. The 
uncertainties are 
influenced by the limited statistical accuracy of the data samples, in 
particular for the bins with larger $log_{10}N_{\mu}^{tr}$ values i.e. primary
energies. 
Nevertheless it is obvious that mass compositions of measured KASCADE samples
differ for different $R_\mu$ bins, since the observation conditions lead to 
mass dependent differences in the observation efficiency at different $R_\mu$, 
leading to distortions of the deduced primary mass composition. This feature 
should be removed after applying a correct efficiency correction procedure. That
necessarily implies again the use of the particular hadronic interaction model.

\subsection{Reconstruction of the primary mass composition}
Efficiency correction factors ($C_{H}$, $C_{O}$, $C_{Fe}$) have been calculated 
in order to adjust the mass composition of measured KASCADE samples 
($P_{H}$, $P_{O}$, $P_{Fe}$) to the primary mass composition 
($P_{H}^*$, $P_{O}^*$, $P_{Fe}^*$) according to the relation \cite{2bade01}:
\begin{eqnarray}
P_H^*:P_O^*:P_{Fe}^*=\frac{P_H}{C_H}:\frac{P_O}{C_O}:\frac{P_{Fe}}{C_{Fe}}\,\,.
\label{h*o*fe}
\end{eqnarray}
As a first step, the simulated spectra given by eq.~\ref{ecudn} have been
normalised to the same value (=1000) for each type of primaries $A$ (H, O, Fe), 
and for all simulated cases of $E_0$, $\theta$ and core position ranges.  
For a given primary $A$, the detected spectra will appear distorted
at the ground level due to the absorption in the atmosphere and the selection
cuts. Only a fraction of the original events are detected by the KASCADE 
detector and reconstructed successfully. The major influence on the values of 
the efficiency correction factors arises from the applied cuts on the 
reconstructed shower events. The correction factors depend on the $R_\mu$, 
$\theta$ and $log_{10}N_{\mu}^{tr}$ ranges, on the multiplicity 
threshold for the \mbox{2.4 GeV} muons and, of course, on the primary type.
For a given primary $A$, in a certain 
$\{R_\mu , R_\mu + \Delta R_\mu ;\,\, 
\theta , \theta + \Delta\theta ;\,\, log_{10}N_{\mu}^{tr} , 
log_{10}N_{\mu}^{tr} + \Delta log_{10}N_{\mu}^{tr} ;\,\, n , n+\Delta n\}$ 
(multidimensional) bin, the correction factor ($C_A$) is given by the sum of
the weights ($w_i$) of the $p$ simulated events which are accepted showers in 
the multidimensional bin:
\begin{eqnarray}
C_A=\sum\limits_{i=1}^p w_i\,\,.
\end{eqnarray}
The weights $w_i$ are determined via the normalised mass spectra 
(eq.~\ref{ecudn}) which get filtered by the observation conditions.
Considering the equivalent number of events 
\begin{figure}[t]
\begin{center}
\epsfig{file=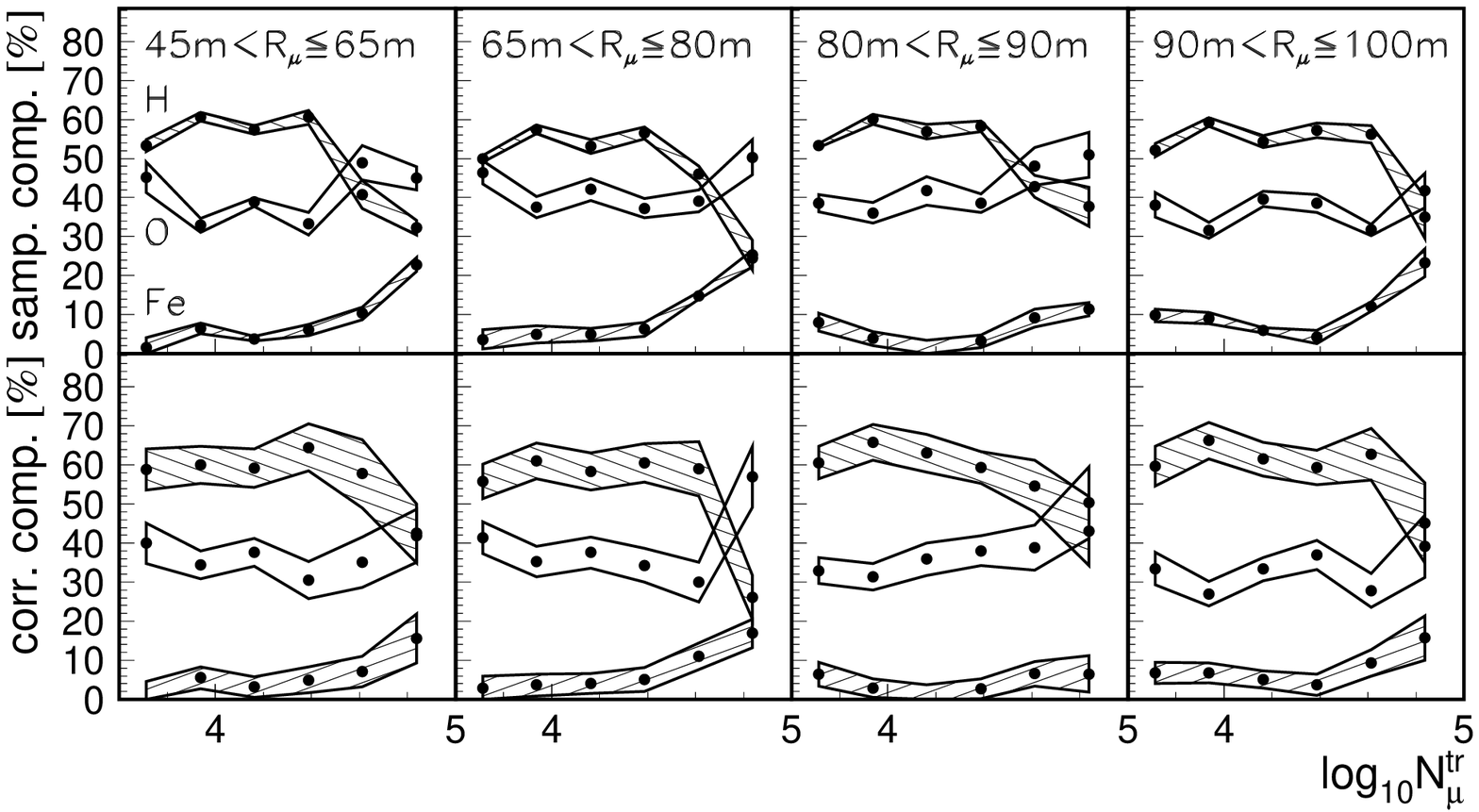, width=13.8cm}
\caption{Variation of the mass composition of measured KASCADE samples 
with $log_{10}N_\mu^{tr}$ estimated from the 
$\{N_{\mu}^{tr},N_e,\Delta\tau_{0.50}^*,\Delta\tau_{0.75}^*\}$ correlation and 
the primary mass composition i.e. before (samp. comp.) and after 
(corr. comp.) efficiency correction shown for different ranges 
$R_\mu$ from the shower centre, for the zenith angle range $0^\circ \leq \theta 
\leq 24^\circ$.}
\label{fig5}
\epsfig{file=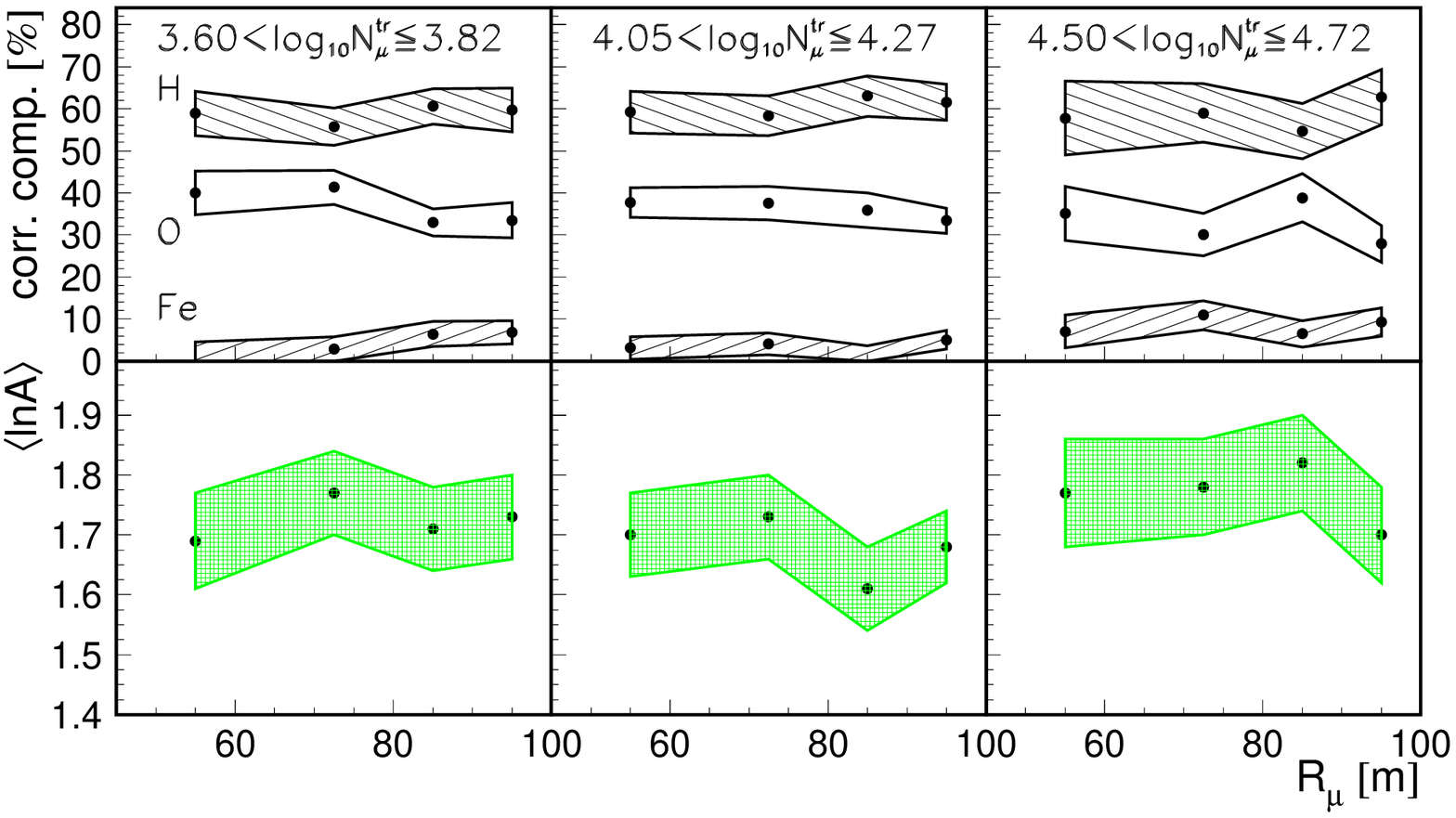, width=13.8cm}
\caption{Variation of the efficiency corrected mass composition and of 
$\langle lnA \rangle$ with $R_\mu$ 
for various $log_{10}N_{\mu}^{tr}$ ranges; $0^\circ \leq \theta \leq 24^\circ$.}
\label{fig6}
\end{center}
\end{figure}
$q$ \cite{CERN93} given by:
\begin{eqnarray}
q=\frac{\left( \sum\limits_{i=1}^p w_i \right)^2}{\sum\limits_{i=1}^p
w_i^2}\,\,,
\end{eqnarray}
the statistical uncertainty (error of the mean value) has been calculated for 
$C_A$ by:
\begin{eqnarray}
\delta C_A=C_A\cdot \frac{\sqrt{q}}{q} =\frac{C_A}{\sqrt{q}}\,\,.
\label{dCA}
\end{eqnarray}
Fig.~\ref{fig5} displays for different $R_\mu$-ranges the
\begin{figure}[b]
\begin{center}
\epsfig{file=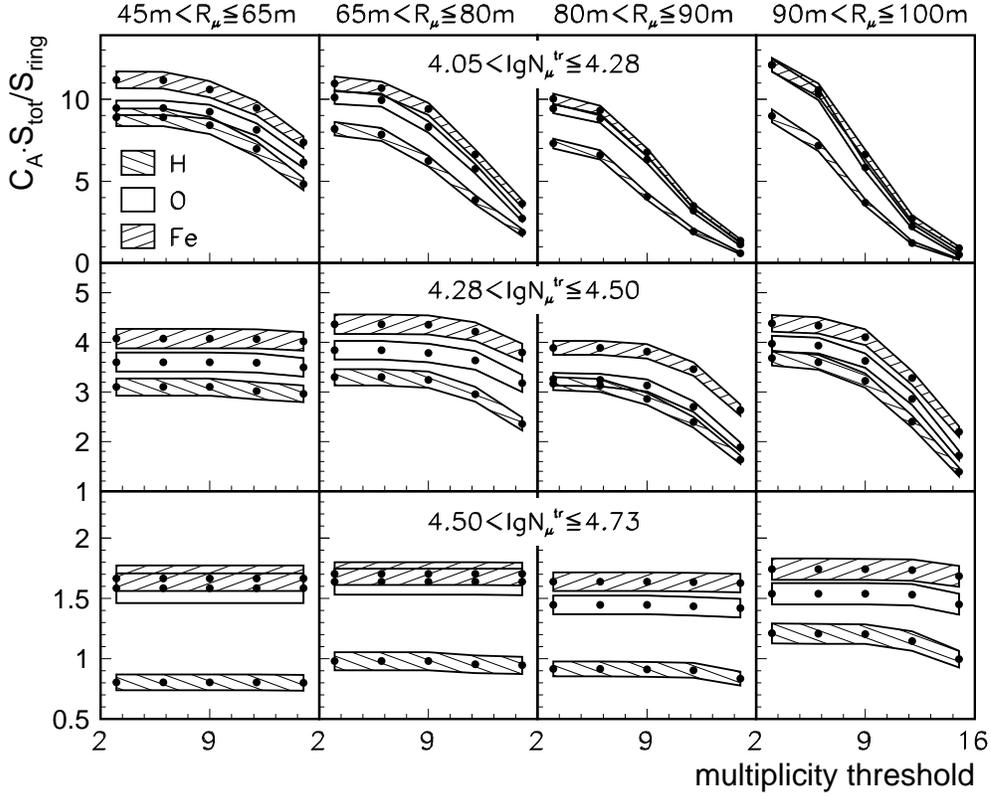, width=14.0cm}
\caption{Variation of the calculated acceptance correction factors with $n_{th}$
for four $R_\mu$ ranges and three $log_{10}N_{\mu}^{tr}$ ranges in the zenithal 
angle range of $0^\circ \leq \theta \leq 24^\circ$. 
The factor $S_{tot}/ S_{ring}$ ($S_{tot}=210\times 210$ m$^2$) accounts for the 
different geometrical areas for different $R_\mu$ ranges.}
\label{fig7}
\end{center}
\end{figure} 
$log_{10}N_\mu^{tr}$-variation of the mass composition of the measured KASCADE 
samples extracted from the 
$\{N_{\mu}^{tr}, N_e,$ $\Delta\tau_{0.50}^*, \Delta\tau_{0.75}^*\}$ correlation 
and the primary mass composition resulting after the correction concerning 
the biased acceptance by the specific observational conditions. 
In the lower part of Fig.~\ref{fig6} the results are shown as variation of 
$\langle lnA\rangle$ 
with $R_\mu$ for different $log_{10}N_\mu^{tr}$ ranges. The uncertainties of the
correction factors (eq.~\ref{dCA}) have been included by propagation via
eq.~\ref{h*o*fe}. It should be stressed that the displayed error bands do not
include the uncertainty of the adopted spectral indices $\gamma$.
Within the overall uncertainties the results obtained for 
different $R_\mu$ ranges are in fair agreement. This can be considered as a 
proof of the consistency of the Monte Carlo simulations invoked for the analysis
of the data.
The correction factors $C_A$ give the fraction of the primary energy spectrum of
the considered primary mass which contributes to the $\{ N_\mu^{tr}, N_e, ...\}$ 
sample observed with conditions specified by the energy threshold of the 
registered muons $E_{th}$, the distance from the shower centre and the 
multiplicity
threshold $n_{th}$. In order to give an impression on these factors, they are 
shown for various cases with their variation with $n_{th}$ in Fig.~\ref{fig7}. 
The variation of $C_A$ is more pronounced at smaller $N_{\mu}^{tr}$ values 
(i.e. primary energies), and as expected from the muon lateral
distributions the acceptance of Fe events in the observation sample is higher 
than for proton events.

\section{Test of the consistency of Monte Carlo simulations}
\begin{figure}[t]
\begin{center}
\epsfig{file=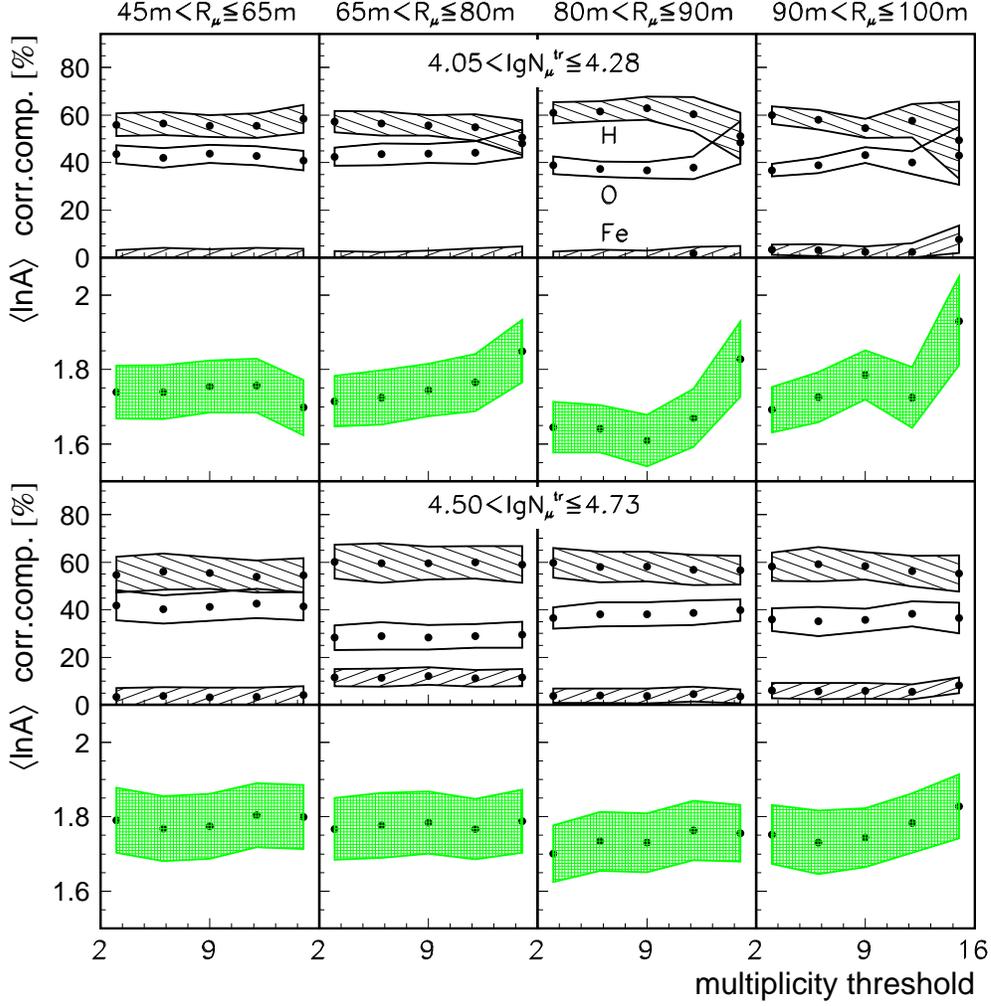, width=14.0cm}
\caption{Variation of the deduced primary mass composition with the multiplicity
threshold $n_{th}$ using the $\{N_{\mu}^{tr},N_e\}$ correlation for the 
non-parametric classification procedure of the events observed at different 
distances from the shower axis. Shown are results for two different 
$log_{10}N_{\mu}^{tr}$ ranges (corresponding to primary energies below and above
the knee in the primary cosmic ray spectrum).
}
\label{fig8}
\end{center}
\end{figure}

As emphasised above, the event samples registered for these measurements are 
particular selections of all EAS events, determined by the observation 
conditions, especially by the energy threshold of the detected muons and the 
multiplicity threshold. Thus the event selections are affected by the lateral 
variation and mass dependence of the muon energy spectrum of EAS. The mass
composition of these samples, as inferred from the correlations of the observed 
EAS observables, are altered as compared to the primary composition, in a way 
varying with the distance $R_\mu$ from shower centre. The acceptance or 
efficiency factors $C_A$ for correcting this effect depend on the Monte 
Carlo simulations and their ingredients like the model descriptions of the 
hadronic interaction. Since the measurements are performed at different 
distances from the EAS centre, the variation of the results after applying the 
$C_A$ factors implies a consistency test of the Monte Carlo simulations since 
the final result should be independent from $R_\mu$ within the given 
uncertainties. This test can be refined and be more stringent when not only the 
variation with $R_\mu$, but also the variation with the multiplicity threshold 
$n_{th}$ is scrutinised.
In Fig.~\ref{fig8} the results of such a refined test are shown, based on the 
dominant $\{N_{\mu}^{tr},N_e\}$ correlation within the observed samples. 
Observables describing the (local) EAS time structure (which are shown to be 
of minor sensitivity) have not been included in the classification procedure
though their measurement defines the test sample. 
The results have been obtained by repeating the full data analysis, described in
the previous sections, for different multiplicity thresholds. In view of the 
rather specific observation conditions, sensitive to variations of the 
EAS muon energy spectrum, we consider the results, even if displaying larger 
uncertainties, as a remarkable confirmation of the consistency of the Monte 
Carlo simulations performed with the CORSIKA code using the QGSJET model.
The test could be improved when samples resulting from specific cuts would be 
analysed by simultaneously classifying the primary energy and the mass, using 
efficient pattern recognition methods as applied in other KASCADE studies 
\cite{roth01,2anto01}. The mass composition resulting from the present studies
is in fair agreement with the results of those investigations. In particular it
corroborates the finding of an increase of $\langle lnA \rangle$ beyond the
knee, i.e. for $log_{10}N_{\mu}^{tr}>$ ca. 4.2.

\section{Conclusions}
The present investigation considers experimental data of the KASCADE experiment 
which are based on observations of local muon arrival time distributions, 
representing the variation of the EAS disc thickness with the distance from the 
shower axis. The first focus of the studies is the question to which extent 
these time quantities and their correlations with other EAS observables 
can help to improve the mass discrimination in the event samples observed with 
realistic experimental, i.e. with KASCADE conditions. Methodically the 
event-by-event analyses apply a non-parametric approach with reference 
patterns provided by CORSIKA simulations of the EAS development, using as 
generators the QGSJET model for the high energy interaction part and 
GHEISHA for the low energy part. It turns out that similarly to other EAS 
observables, the local time quantities provide only a marginal contribution to 
the mass discrimination as compared to the dominant 
$\{N_{\mu}^{tr},N_e\}$ correlation, at least in the relatively limited range 
of \mbox{$R_\mu <100$ m}. It should be emphasised that this statement does by 
no means disqualify the sensitivity of global muon arrival time distributions 
measured relative to the arrival time of the shower centre. 
In fact, 
studies including the curvature of the shower front indicate a considerable 
enhancement of the mass discrimination, in particular at larger distances from 
the shower core and higher primary energies \cite{2bade01,bran01}.\\
The event sample collected with the observation of muon arrival time 
distributions in the KASCADE experiment is a subset of all EAS events measured 
with $100\%$ efficiency for $log_{10}N_{\mu}^{tr}\geq$ approx. 3.6. As 
consequence 
of the muon energy threshold \mbox{$E_{th} = 2.4$ GeV} and the condition 
of a registered multiplicity $n\geq 3$ the original primary mass composition 
in the registered samples gets distorted in a way which is sensitive to lateral 
variations of the integral EAS muon energy spectrum. In order to determine the 
fraction of all shower events which are accepted in the specific subset, 
specified by the measuring conditions, Monte Carlo simulations have to be 
invoked for the calculation of corresponding acceptance or efficiency factors. 
The variation of such corrections with the distance of observation $R_\mu$ 
from the shower axis and with the multiplicity threshold $n_{th}$  
provides the possibility to test the consistency of the Monte Carlo simulations 
with the data. Even admitting the large uncertainties due to the limited number 
of events observed and simulated, our results indicate a remarkable consistency 
of the performed Monte Carlo simulations using the QGSJET model as generator.

{\ack
We acknowledge various useful discussions with Prof. Dr. M. Petrovici about the 
aspects of the results elaborated in this paper. The authors would like to thank
the members of the engineering and technical staff of the KASCADE collaboration 
who considerably contributed with enthusiasm and expert engagement to the 
success of the reported measurements. The work has been supported by the 
Ministry of Research of the Federal Republic of Germany. 
The Romanian Ministry of Education and Research provided a grant for supporting 
the collaborating group of the National Institute for Physics and Nuclear 
Engineering of Bucharest. The Polish collaborating group 
of the Cosmic Ray Division of the Soltan Institute for Nuclear Studies, Lodz,
is supported by the Polish State Committee for Scientific Research (grant No. 
5 P03B 133 20). The KASCADE collaboration work is embedded in the frame of 
scientific-technical co-operation (WTZ) projects between Germany and Romania 
(ROM 99/005), Poland (POL 99/005) and Armenia (ARM 98/002).
}

\end{document}